\documentclass[10pt,journal,compsoc]{IEEEtran}

\usepackage{graphicx, subfigure}
\usepackage{amsmath}  
\usepackage{array}
\usepackage{float}
\usepackage{comment}
\newtheorem{theorem}{Theorem}
\newtheorem{lemma}[theorem]{Lemma}
\newtheorem{proof}[theorem]{Proof}
\begin{document}
%
\title{Shifting Behaviour of Users: Towards Understanding the Fundamental Law of Social Networks}
%
%
%
%

\author{\IEEEauthorblockN{Yayati Gupta\IEEEauthorrefmark{1},
Jaspal Singh Saini\IEEEauthorrefmark{2},
Nidhi Sridhar\IEEEauthorrefmark{3} and
S.R.S. Iyengar\IEEEauthorrefmark{4}}\\
\IEEEauthorblockA{\IEEEauthorrefmark{1}\IEEEauthorrefmark{2}\IEEEauthorrefmark{4}Department of Computer Science,
Indian Institute of Technology, Ropar\\
Email: yayati.gupta@iitrpr.ac.in\IEEEauthorrefmark{1}, jaspal.singh@iitrpr.ac.in\IEEEauthorrefmark{2}, sudarshan@iitrpr.ac.in\IEEEauthorrefmark{4}}

\IEEEauthorblockA{\IEEEauthorrefmark{3}Department of Information Technology,
National Institute of Technology Karnataka, Surathkal\\
Email: nidhisridhar16@gmail.com\IEEEauthorrefmark{3}}
}

\IEEEtitleabstractindextext{%
\begin{abstract}
Social Networking Sites (SNSs) are powerful marketing and communication tools. There are hundreds of SNSs that have entered and exited the market over time. The coexistence of multiple SNSs is a rarely observed phenomenon. Most coexisting SNSs either serve different purposes for its users or have cultural differences among them. The introduction of a new SNS with a better set of features can lead to the demise of an existing SNS, as observed in the transition from Orkut to Facebook. The paper proposes a model for analyzing the transition of users from one SNS to another, when a new SNS is introduced in the system. The game theoretic model proposed considers two major factors in determining the success of a new SNS. The first being time that an old SNS gets to stabilise. We study whether the time that a SNS like Facebook received to monopolize its reach had a distinguishable effect. The second factor is the set of features showcased by the new SNS. The results of the model are also experimentally verified with data collected by means of a survey.
\end{abstract}

}

\maketitle

\IEEEdisplaynontitleabstractindextext

%
\IEEEpeerreviewmaketitle

\ifCLASSOPTIONcompsoc
\IEEEraisesectionheading{\section{Introduction}\label{sec:introduction}}
\else
\section{Introduction}
\label{sec:introduction}
\fi

%
%
%
%

\IEEEPARstart{T}{he} future of Social Networking Sites stands as a highly debated topic today. Since antiquity, humans have found various means for socializing and communicating with fellow humans. The communication systems have evolved from pigeons, telegraph, light signals and telephones to Social Networking Sites(SNSs), which today stand as a billion dollar industry. SNSs today are dominating the web and increasingly becoming objects of scholarly research. Their history is a long one \cite{ellison2007social}. The first major SNS to hit the internet was SixDegrees, launched in May 1996. It was followed by a number of different sites including Napster(1999), Friendster(2002) and MySpace(2003). In 2004, Orkut and Facebook were launched in succesion. The initial brisk growth of Orkut  died out after the launch of Facebook. Orkut faded, though other SNSs like LinkedIn, Twitter, etc. continued to grow parallely not only in terms of number but also in terms of user activity. Google Plus strived to stand collaterally with Facebook, but it failed to attract the monopolized market of Facebook. The activity of people on Google Plus remained low. Generally, it is difficult for two SNSs to coexist together in the same internet space until they have very different aims. Hence, most of the time, it is only one SNS that rules the web. We term this phenomenon of monopolization as the ``Fundamental Law of Social Networks''. Today, Facebook has largely become synonymous with socializing online. Popular debates on- ``Can Facebook ever be overthrown?'' have not been able to provide convincing evidence to support for or against the argument. According to a study by Princeton researchers, Facebook is predicted to lose 80\% of its users by the year 2017 \cite{cannarella2014epidemiological}. The researchers claimed that Facebook has already reached its peak popularity phase and may not even exist by 2021. Looking at the history of the SNSs and the ongoing debates, the future of social media is worth speculating.\\

In this paper, we address the very question of whether a new SNS can make a mark, when the existing market has been monopolized for a long period of time. SNSs can also be considered as a type of products which keep entering and leaving the market. Although, the adoption of a product by people and its cascading pattern in the population has been studied \cite{adar2004implicit, adar2005tracking, saxena2015understanding}, the switching of a user to a new SNS has not been looked at in detail. Unlike the case of traditional commodities, it is useless for a user to shift to a SNS if none/very few of his friends are present on the new SNS. The entire concept of a SNS is based on interpersonal connections, and hence it demands a new study altogether. Shifting to a new SNS requires the additional cost of creating new contacts and a loss of old connections. So, people tend to be careful while shifting completely to a new SNS. We propose a model to simulate the future of a SNS when a new SNS arrives in the market. Moreover, we employ a questionnaire survey to determine the parameters required for the simulation of the model.\\

We model a SNS as a set of features. We consider the users' time to be distributed across these features. According to our model, a user's decision to shift to the new SNS is based on the two opposing factors :
1. Supporting shift: Novel features encompassed by new SNS that could compel the users to spend less time on the old SNS.
2. Opposing shift: The attachment factor associated with the old SNS that denotes the comfort level of the users with the old SNS. This attachment factor increases with time.
Our model simulates the transition phase of a SNS i.e. the time period during which the new SNS arrives. It takes into consideration both the diffusive and non-diffusive shifts possible. We show that the presence of at least a few new features in the new SNS is a necessary condition for it to succeed. The paper further provides an insight as to why the efforts made by Google Plus to reach Facebook’s popularity were unfruitful, whereas Facebook was outstandingly successful in overtaking Orkut.\\

The paper is organised as follows: First, we highlight important work done in this direction in Section 2 which is followed by basics in Section 3. The proposed model is discussed in Section 4. Section 5 gives an overview of the survey conducted. The paper continues with the Results and Simulations in Section 6 and is concluded along with future work in Section 7.


\section{Related Work}
Our work derives its theme from two lines of research in Social Network Analysis: 1) Study of the cascading pattern on SNSs and 2)  Temporal evolution of Social Networking Sites.\\

One of the most widely studied problem in Social Network Analysis is understanding the cascading pattern of an idea/information on an online social network, generally termed as a meme \cite{adar2004implicit,adar2005tracking,saxena2015understanding}. People tend to share memes on SNS because of interest or altruism. The spread of such memes on a network has been studied extensively. Such kinds of study apply to the market products also. Many researchers have tried predicting the virality of a meme or a product by analyzing its content and the initial pattern of spreading in a  SNS \cite{berger2012makes,berger2010social,weng2013virality}. One of the most important applications of SNSs lies in digital advertising \cite{leskovec2007dynamics}. Today, social networking sites turn out to be the most widely used platforms for the advertisement of a wide variety of products. Most of the population is online and hence triggering a group of people to adopt a product may result in the product being globally popular due to a large cascade. Kleinberg et al. gave an approximation algorithm for choosing the right set of seed nodes that results in the largest cascade over a given network \cite{Kempe:2003:MSI:956750.956769}. Ugander et al. observed the cascading process at a local level. He extrapolated the linear threshold model theory to include the diversity of a node's neighbours in addition to their number.  Here, the diversity  is quantified by the number of connected components in the induced subgraph on its neighbours \cite{Ugander17042012}.
Watts conducted an experiment \cite{salganik2006experimental}  to prove the increase in the  unpredictability of the success of a product in an environment with increasing social influence.  Another similar paradox, that of a globally sparse phenomenon giving you an illusion of being globally viral was discussed in \cite{lerman2015majority}.\\

Extensive research has been done towards understanding the growth of online social networks. Leskovec et al. proposed a model for the evolution of a social network over time\cite{leskovec2008microscopic}. This model is based on the data collected from different social networking sites like Flickr, LinkedIn, Delicious and Answers. Kumar et al. explained the process of a SNS evolution by categorising  the nodes of the network into 3 types \cite{kumar2010structure} : Passive, Linker and Inviter. Mislove et al. observed a proximity bias in new links that are being created\cite{mislove2008growth}. Kairam et al.  \cite{Kairam:2012:LDO:2124295.2124374}  have studied the life and death of online groups in different Ning communities. They proposed two ways for the growth of a new group: Diffusion and Non-Diffusion. Anderson et al. observed homophily in cascading invitations after analysis of LinkedIn’s sign up cascades \cite{Anderson:2015:GDV:2736277.2741672}.\\

Benevenuto et al.\cite{benevenuto2009characterizing} experimentally analyzed user behaviour on different social networking sites and observed that people used to spend maximum amount of time on Orkut. Wilson et al.\cite{wilson2009user} and Viswanath et al.\cite{viswanath2009evolution} emphasized the importance of the level of interaction between 2 nodes in a social graph. They also studied the evolution of these interactions with time on the Facebook network.\\

\cite{ribeiro2015modeling} have modelled the shift based on Daily Active Users(DAU) of MySpace and Facebook through catalytic conversions. We present a simple game theoretic model that demonstrates transition/no transition of people from one social networking website to another based on a number of other parameters like time, etc.

\section{Basics}
\begin{figure}[h!]
\begin{center}
\includegraphics[width=6cm]{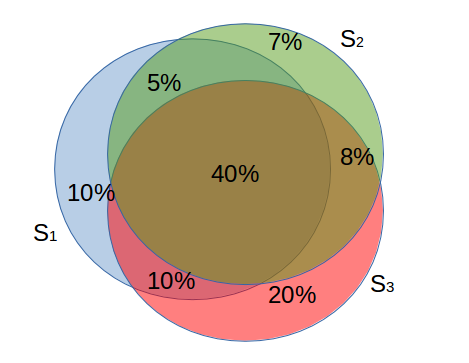}
 \caption{An example of feature space}
 \end{center}
\label{fig1}
\end{figure}

Each SNS has an underlying implicit network, with nodes of the graph being the users of the SNS and the links between them representing a type of connection between them. For example, on Facebook, the connections represent friendship links whereas on Twitter it represents follower-followee relationship.\\
Each user has only certain stipulated amount of time which she spends on SNSs. We define $\Delta(u)$ to signify this \emph{time span} of a user $u$. We further observe with the help of a survey conducted, that the variance of time span across different users of a SNS is low. Therefore, for the sake of simplicity, we assume that each user spends nearly equal amount of time on SNSs i.e.  $\Delta(u) = \Delta$ for all users $u$. \\
Modeling the shifting phenomenon of users from one SNS to another requires us to concretely define a \emph{social networking site} (SNS). There have been earlier attempts to define a SNS \cite{ellison2007social}, but for the ease of modeling, we adopt a slightly modified definition compared to those already proposed. \\
A \emph{social networking site}($S$) act as a platform for people sharing common interests, beliefs, ideas and hobbies to communicate and socialise with each other. In today\rq s world, a SNS provides us an opportunity to constantly stay connected with our friends, majorly though \emph{passive engagement} \cite{easley2010networks}\cite{benevenuto2009characterizing}. In addition, a SNS can be characterized by the following attributes:
\begin{itemize}
\item Set of all features available ($F(S)$)
\item User-Interface of the website (excluding the UI of features)
\item Latency
\end{itemize}
A \emph{feature} ($f$) on a SNS $S$ is identified by its functionality, volatility, UI, etc. A feature $f$ on a SNS is considered to be important if it consumes a large amount of time of its users. To quantify the same, we introduce a new parameter $\phi(f,u)$ for a feature $f$ and a user $u$, which is defined as the fraction of the user $u$'s timespan that is spent on feature $f$ available on the SNS $S$.\\
Since most of the users on a social networking site have a large number of friends, it is impossible for a user to explicitly devote time for each of her friends. Therefore, features that allow for passive engagement generally tend to be more popular. Most features act as a platform for all users to participate as a group and are not based on one-to-one communication. Therefore, a user\rq s time span is distributed across the features of SNSs and not across her friends. This claim can be supported by the clickstream analysis done in \cite{benevenuto2009characterizing}, where a large percent of net time spent by users is spent on features supporting passive engagement.\\
For a better visualization of how the feature sets of different SNSs intersect, we introduce the concept of a feature space (FS). We introduce it with the help of an example, shown in Figure \ref{fig1}. Let $u$ represent a individual who uses all three SNSs $S_1$, $S_2$ and $S_3$ in the system, then\\
$x_1$ = $10\%$ =  percent of time span $\Delta$ spent by the user $u$ on the features which are available only on $S_1$	\\
$x_{2,3}$ = $8\%$ = percent of time span $\Delta$ spent by the user $u$ on features which are present precisely on both $S_2$ and $S_3$	\\
$x_{1,2,3}$ = $40\%$ = percent of time span $\Delta$ spent by the user $u$ on features which are present on all three SNSs. \\
Although feature space is an innovative approach of understanding the interplay of feature sets across different SNSs, it does not provide complete information about the system under observation. For example, from figure \cite{}, $10\%$ of time span $\Delta$ of the user $u$ is spent on features which are present precisely on both $S_1$ and $S_3$. But using just the $FS$ one cannot comment on the fraction of this $10\%$ time span that is spent on $S_1$ and $S_3$ individually. A formalization of the concept of a feature space can be made, although not required for this work.

\section{Model}
We analyze a system consisting of $n$ users, where only one SNS $S_1$ is present. Therefore, each of the $n$ users spend their entire time span $\Delta$ on $S_1$. Let $G_1$ represent the underlying social network on SNS $S_1$. Now, let a new SNS $S_2$ be introduced into the system. \\

When the new SNS $S_2$ is launched, some individuals tend to use it because of an internal urge or due to an explicit incentive. Even if an individual starts using the SNS $S_2$, she do not stop using the SNS $S_1$ completely. This is largely attributed to the attachment and trust factor  associated with the old site $S_1$ \cite{benevenuto2009characterizing}. It is also illogical to betray one\rq s social networking site altogether and try another one, since the user will be risking her means to socialize online.  We assume that a major driving force for a user to start using the new SNS $S_2$ is the new set of features that $S_2$ offers. We further assume that during the transition phase i.e.\ around the time when $S_2$ was introduced, the common features are used by individuals only on $S_1$. Although, people may gradually start using the common features on $S_2$ later on in time. All the assumptions mentioned above are supported by the survey results discussed in section 5. \\

To understand the cascading phenomenon from one SNS to another, one needs to have a microscopic view of the underlying graph $G_1$. We look at an exclusive user $u$ of $S_1$ and examine the reasons which can incentivize her to use $S_2$ as well. Similar to the concept given in \cite{Kairam:2012:LDO:2124295.2124374}, we propose two mechanisms through which a user can be encouraged to use $S_2$.

\begin{figure}[h!]
    \centering
       \includegraphics[width=8cm]{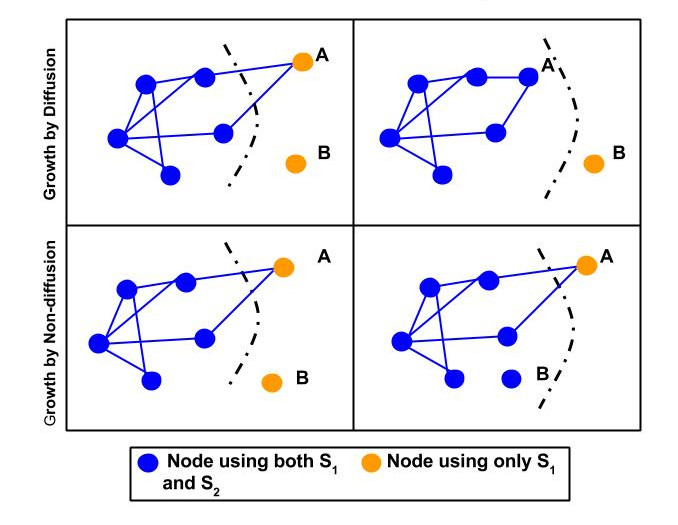}
       \caption{Diffusive shift vs Non-Diffusive shift}
       \label{figure1}
\end{figure}

\begin{enumerate}
\item \textbf{Non-Diffusive Shift} is the process by which a user shifts from one social networking site to another as a result of marketing strategies used by the SNS $S_2$. To account for this in our model, we associate a  small probability $p$ with which an exclusive user of $S_1$ starts using $S_2$ at a given point in time. We term $p$ as the non-diffusive probability per unit day.

\item \textbf{Diffusive Shift} can be defined as the process by which a user starts using the new social networking site $S_2$ due to influence from her friends, mainly by the principles of social reinforcement and homophily. 
Diffusive shift can be catalyzed using various methods, few of which are mentioned below:
\vspace{-1cm}
\begin{itemize}
\item Reduction in the  time spent by a user $u$'s neighbors on $S_1$
\item Invitations to join $S_2$ from neighbors of $u$ who are present on $S_2$
\item Word of mouth advertisements
\end{itemize}
\vspace{-0.8cm}
Figure \ref{figure1} illustrates both these phenomena.
\end{enumerate} \vspace{-0.1cm}
\vspace{-0.5cm}
Consider a user ($u$) who is an exclusive user of $S_1$ having degree $d$. Out of these, $d_1$ neighbours are exclusive users of $S_1$ whereas $d_2$ of them use both $S_1$ and $S_2$ simultaneously. Let $T_1$ and $T_2$ represent the net time spent by the neighbours of user $u$ on SNS $S_1$ and $S_2$ respectively.

\begin{figure}[H]
\begin{center}
\includegraphics[width = 8cm]{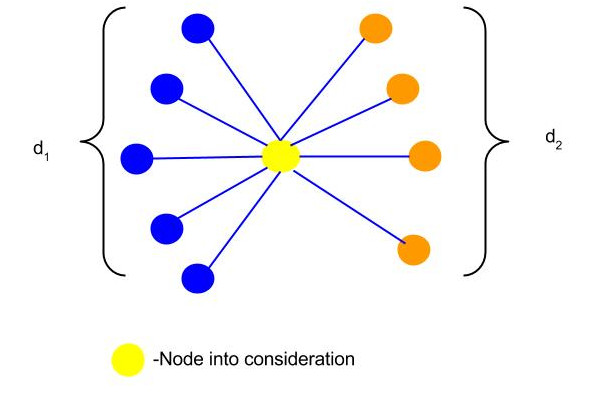}
\end{center}
\caption{Neighborhood of the node under consideration for diffusive shift}
\end{figure}

\begin{lemma}
$T_1 = \Delta(d - d_2x_2)$

\end{lemma}

\begin{proof}
$d_1$ friends of user $u$ spend their entire time span $\Delta$ on $S_1$. From the definition of a Feature Space and the assumption involved that common features are used by the users on $S_1$, $d_2$ friends of $u$ spend $(1-x_2)$ fraction of their time span on S1. Therefore,
\begin{align*}
 T_1     &= \Delta(d_1) +\Delta(1-x_2)(d_2)  \\
    &= \Delta(d - d_2) + \Delta(1-x_2)(d_2)  \\
    &= \Delta(d - d_2 +d_2- x_2d_2 )  \\
    &= \Delta(d - d_2x_2)
\end{align*}

\end{proof}

\begin{lemma}
$ T_2 =  \Delta d_2x_2$
\end{lemma}
\begin{proof}
Follows from Lemma 1
\end{proof}

Most of the features of a SNS are valued only because of the presence of a large pool of users who use it, for example: NewsFeed, Groups,etc. Hence, even if a feature $f$ may have attracted user $u$ initially with its novelty and UI, for it to be a feature where the user $u$ spend time regularly, the feature must also be used by the friends of $u$. Therefore, if most of user $u$\rq s friends start using $S_2$, $T_1$ would reduce, and at some stage it will incentivize user $u$ to start using $S_2$ as well. We introduce a new parameter $\epsilon$ to determine the threshold for diffusive shift. A user $u$ starts using $S_2$ iff $T_2 > \epsilon(T_1+T_2)$ i.e.\ once the time spent by friends of $u$ on $S_1$ fall below a fraction of their net time span on SNSs, user $u$ starts using $S_2$. This constant $\epsilon$ is called the attachment factor of SNS $S_1$. The attachment factor always lies between $0$ and $1$. It accounts for the attachment that the users develop towards a SNS over time. Lower is the value of $\epsilon$, easier is it for the cascade to occur, and higher the value of $\epsilon$, tougher it will be for the diffusive shift. From [1], with time the attachment factor generally increases. A detailed study of the attachment factor is out of the scope of this paper.

\begin{theorem}
If $\epsilon \geq x_2$, diffusive shift can never occur. 
\end{theorem}

\begin{proof}
Consider a user $u$ with $deg(u)$ = $d$ = $d_1 + d_2$ where $d_1$ = number of friends of $u$ who exclusively use $S_1$
\begin{align*}
\Delta d \epsilon &\geq \Delta d x_2        \tag{From the given condition} \\
\Delta d \epsilon  &\geq \Delta d_2 x_2      \tag{Since $d_2 \leq d$} \\
(T_1 + T_2)\epsilon &\geq T_2    \tag{from lemma 1 and 2, $(T_1 + T_2) = \Delta d$}
\end{align*}

\end{proof}
Therefore, the user $u$ can never start using $S_2$ with the help of diffusive shift. Since we picked an arbitrary user $u$, no node in $G_1$ can be diffused to the other SNS.\\
Non-diffusion is generally a less important cascading factor as  compared to diffusion \cite{Kairam:2012:LDO:2124295.2124374}. Therefore, from the previous theorem, If $\epsilon \geq x_2$, a new SNS will never be able to uproot the existing SNS. 
Even if $\epsilon$ and $x_2$ are close by, diffusive shift does not propagate for large distances in the network and hence the SNS $S_1$ would still largely monopolize the market.
Since, with time the attachment factor $\epsilon$ increases, it will keep becoming harder for a newcomer to overtake the current popular SNS. This theorem also gives a lower bound on the novelty ($x_2$) that $S_2$ must possess for it to have any chance of beating the current favourite SNS.

\section{Survey}
We consider as a case study, the shift of users from Orkut to Facebook and Facebook to Google Plus. Due to lack of availability of actual data for verification, we conducted an online survey asking our participants a plethora of questions. The participants belonged to a wide age group. 189 of these participants were between 17 to 25 years of age. Out of the remaining, 27 belonged to the age group 25-40. A very small fraction of participants were between 13 -17 or more than 40 years old. 
The survey comprised of a wide variety of questions ranging from their reason of shifting to other SNSs to the amount of  time they spend on the features of various  SNSs. Other information asked from the participants was:
\begin{itemize}
\item the time they spent on different SNSs
\item the features they look for and use on a SNS
\item the sites they currently use and their distribution of time span across these websites
\item  The reason for their inactivity on Google Plus (Which was observed to be the case with most survey participants)
\end{itemize}
In total, 222 responses were reported. The major observations made from the survey are described below. The full survey details can be found in the appendix section of our arxiv paper\cite{gupta2015shifting}  \\
97 \% of the participants used Facebook, 29.7\% of them used Twitter (mostly along with Facebook) and 32\% of them used Google Plus( mostly along with Facebook). The average time spent on Facebook is 43.8 minutes which is relatively much higher than that spent on Twitter( 8.71 minutes) and Google Plus( 9.66 minutes). This shows that a majority of the people are bound to use the more popular SNSs, both in terms of number of users and also the time spent by the users of the site on it. Both these statistics support our model.
\begin{figure}[h!]
\centering
\includegraphics[width=8 cm]{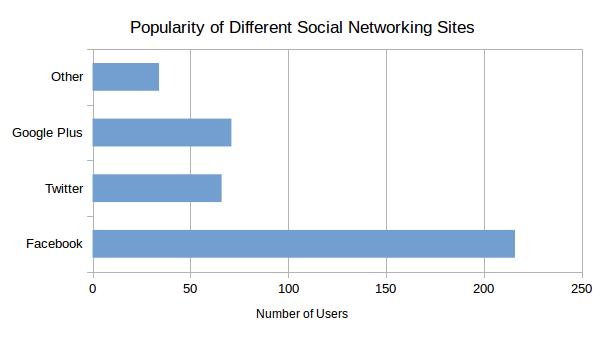}
\caption{Popularity of SNSs}
\label{pop}
\end{figure}
The participants were further asked to rate the features of Facebook, Google Plus and Orkut based on the time spent on it: (5) if they spent most of their time on that SNS using that particular feature and (0) if they don’t use that feature at all. Based on these ratings, we calculated $\phi(f,u^*)$ for each feature $f$, where $u^*$ is a hypothetical user who used all SNSs in the system. Further using these values we developed the feature spaces for both the systems under consideration, as shown in figure\ref{fs1}.

\begin{figure}
\includegraphics[width= 0.23\textwidth]{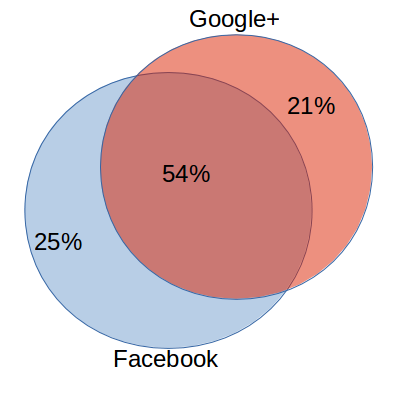}
 \includegraphics[width= 0.25\textwidth]{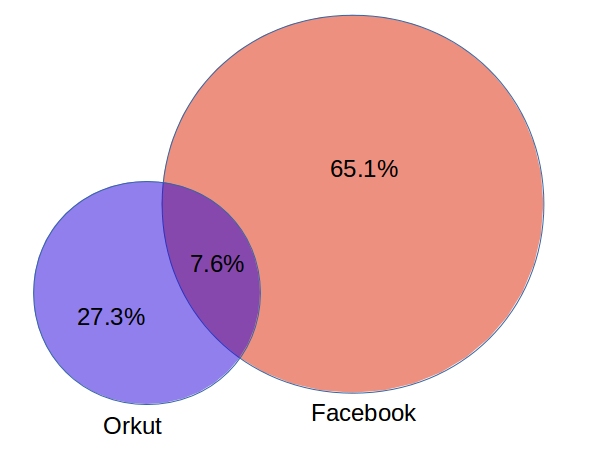}
\caption{ \textit{Left: }Feature Space of Facebook and Google Plus \textit{Right :}Feature space of Facebook and Orkut}
\label{fs1}
\end{figure}

Similarly, the values for Orkut and Facebook were calculated. This is displayed in figure \ref{fs1}.

\section{Results and Simulations}

We  calculated the feature set of Facebook, Google Plus and Orkut by conducting a survey. The  constructed feature spaces are shown in figure \ref{fs1} . Next, we simulate the dynamics of diffusive and non-diffusive shifts on real world graphs. For this purpose, the dataset is obtained from snap \cite{snapnets} and the information obtained from the survey is also employed. \\

\begin{figure}[h!]
    \centering
    \subfigure[$\epsilon=0.5$]
    {
       \includegraphics[width=5.8cm]{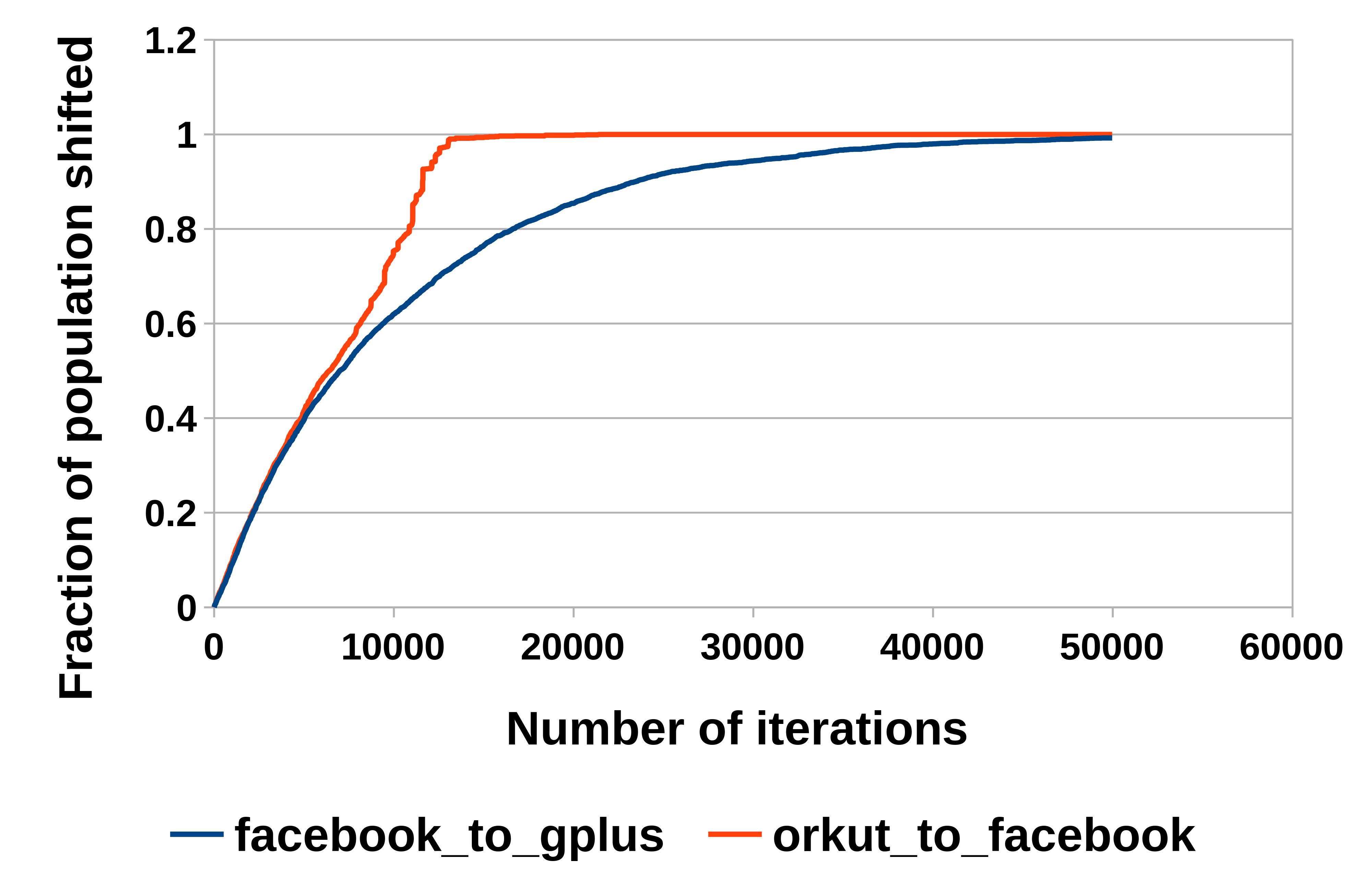}
       
    }
    \subfigure[$\epsilon=0.3$]
    {
       \includegraphics[width=5.8cm]{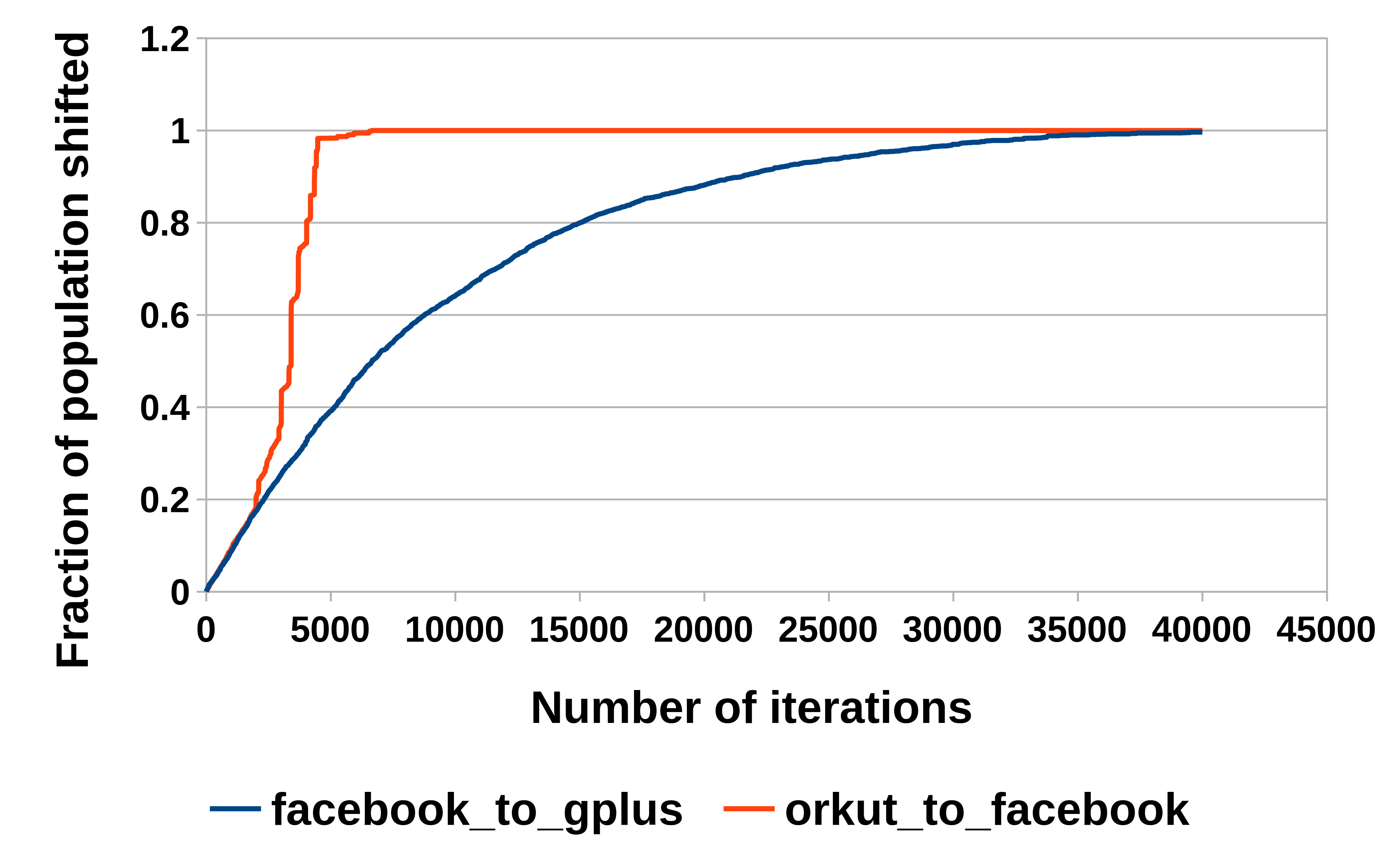}
       
    }
    \subfigure[$\epsilon=0.1$]
    {
       \includegraphics[width=5.8cm]{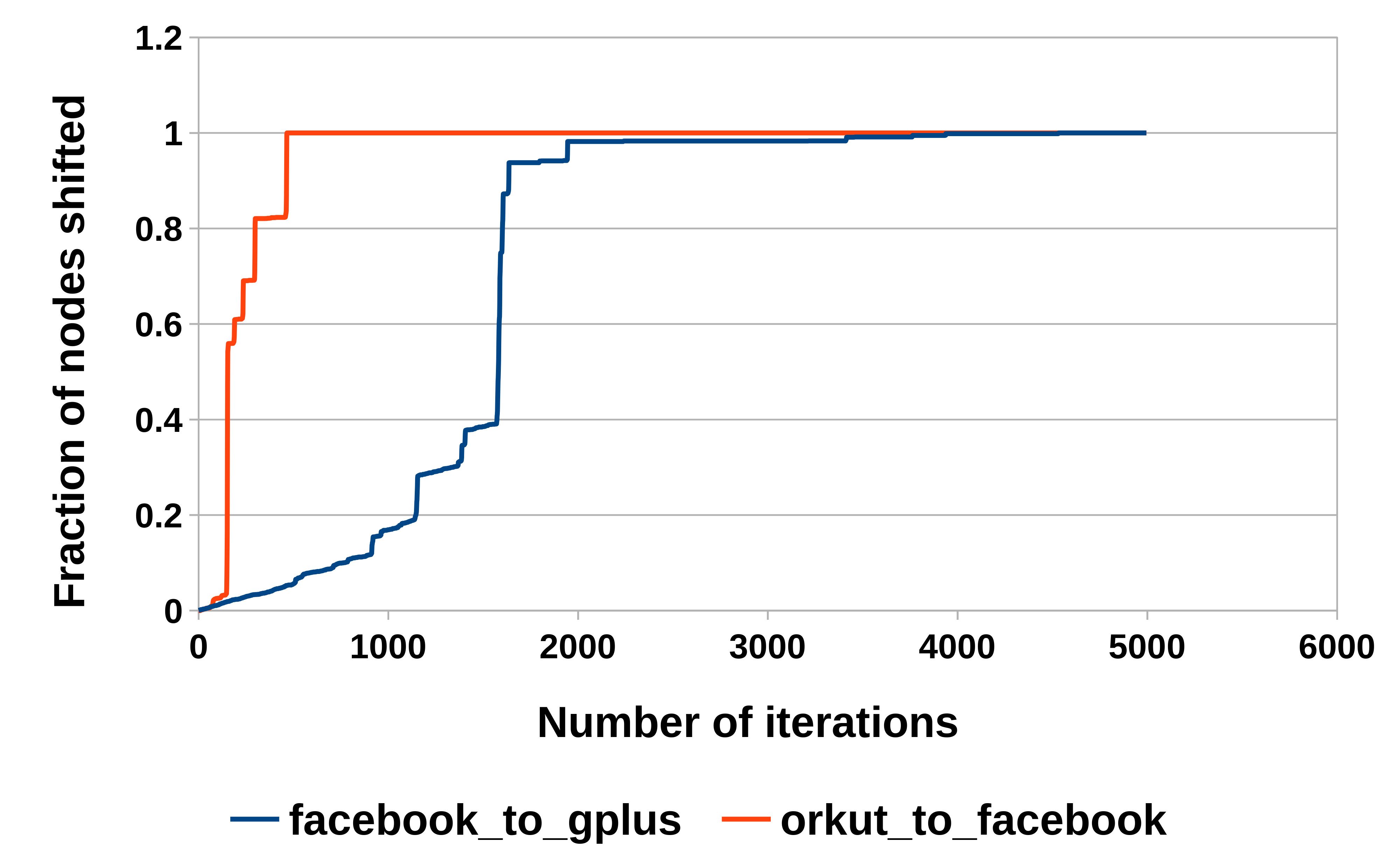}
       
    }
  \caption{Shifting behaviour of nodes}
  \label{shift}
\end{figure}

We perform two simulations: 1) Initially, all nodes are assumed to be on Orkut and then the nodes are shifted to a new SNS Facebook using the proposed model and 2) assume all nodes to be present on Facebook, and then we shift nodes to a new SNS Google Plus. We plot the net cascade of the network to the new SNS as a function of time. We also vary the attachment factor ($\epsilon$) for a better understanding of this parameter. The non-diffusion probability ($p$) is kept as low as $0.001$, since our major focus is to observe the impact of diffusive shift. The results of this simulation is shown in figure \ref{shift}. It can be observed from the plots that as $\epsilon$ increases, the number of time steps required to completely shift the entire population in the network also increases. This is the proof that increasing $\epsilon$ is an indication of the network being less susceptible to change. Also, it can be observed from the plots that, the transition/cascade from Orkut to Facebook was way faster when compared to the percent cascade from Facebook to Google Plus, this observation can be clearly accounted for by looking at the feature spaces of both the systems. Therefore, we can conclude that the two major factors which can drive the success of a new SNS is the amount of novelty it possess and the time at which it was launched (the sooner the better).\\
The results are shown in the figure \ref{shift}.

\section{Conclusion and Future Work}
The paper proposes a game theoretic model to simulate the shifting behaviour of population
across social networking websites.
The key elements of the model considered are the amount of novelty in the
new website and the attachment factor associated with the old website.
Although the major contribution of the model is to understand the transition in SNSs, it can be applied to other scenarios as well. In other transitions, there might not always exist an explicit graph as in the case of a social networking site.\\
In this paper, we characterise a social network as a set of features. We discover that the most used features were those which involved passive interaction of users. Also, our model only considers the progressive
case, wherein a user does not stop using the new site and returns to the
first site again. The non-progressive state will be further explored in future work.\\
Due to lack of experimental data and for the sake of simplicity, we assumed
the time span to be constant for all the individuals in the system, which is
generally not the case. The model can further be improved by considering the
exact distribution of delta across the population. The attachment factor(inertia)
that has been left as a blackbox in this paper, is observed to have a direct
correlation with time spent on a product over time. There may be some individuals
who are relatively new to the SNS compared to others, which implies that the attachment factor varies from individual to individual as well. Most of the SNS’s have a plethora of
small features, which do not attract a significant chunk of the time span of its
users. Their role is vaguely understood, and needs to
be explored in future. \\
Marketing strategies
may help in the initial growth of a product’s popularity through non-
diffusive shift, but the quality of the product is important in making the users
stay for longer periods.  It appears that the success mantra for a new social networking site is to
come up with a lot of new features supporting passive engagement and launch
the website in a window of minimum time. The more time it takes to launch, the tougher it
becomes to capture people$\rq$s attention.``Don$\rq$t Wait. Just Do It'' looks like the most appropriate
way to increase the odds of an idea becoming successful.

\bibliographystyle{IEEEtran}
\bibliography{fund}

\end{document}